# Einstein's aborted attempt at a dynamic steady-state universe.

To appear in the Gedenkband "*In memoriam Hilmar W. Duerbeck*". (Acta Historica Astronomiae)

Harry Nussbaumer
Institute for Astronomy
ETH Zurich
CH-8093 Zurich
Switzerland

## Abstract

In June 1930 Einstein visited Cambridge where he stayed with Eddington who, three months before, had shown that Einstein's supposedly static universe of 1917 was not stable. This forced Einstein to rethink his cosmology. He spent January and February 1931 at Pasadena. There, he discussed cosmology intensely with Tolman, conscious that he had to replace his original model of 1917. However, at the end of February he still had not made up his mind about an alternative. The Albert Einstein Archives of Jerusalem (AEA) hold an undated draft, handwritten by Einstein, which I date to the beginning of January 1931. In this draft Einstein hopes to have found a solution to the cosmological problem: a stationary, dynamic universe in expansion. His model was stationary because particles leaving a given volume were replaced by particles created out of the vacuum, anticipating an idea of Bondi, Gold and Hoyle published in 1948. He saw the cosmological term as energy reservoir. However, he then realised that his calculations contained a numerical error. When the error was corrected his steady-state-model collapsed.

## *Introduction*

Einstein's diary of January 3, 1931, holds a cryptic statement: *Arbeiten am Institut. Zweifel an Richtigkeit von Tolmans Arbeit über kosmologisches Problem. Tolman hat aber Recht behalten.* (Work at the Institute. Doubt about correctness of Tolman's work on the cosmological problem. But Tolman was right.) Einstein had just arrived in Pasadena, where he would be spending January and February before returning to Berlin. Six months before, Eddington had shown that Einstein's static universe of 1917 was unstable. Thus Einstein had to rethink his cosmology, for him that was a tortuous path. In Pasadena, Tolman was Einstein's man for cosmological discussions. In which respect was Tolman right and Einstein wrong? An inconspicuous draft in the Albert Einstein Archives of Jerusalem (AEA) might hold the clue. Einstein had most probably hoped to have found an ingenious way out of the cosmological problem by postulating a steady state universe with everlasting creation of new matter. But it turned out to be a mirage caused by an error in calculation.

## *Einstein visits Eddington*

June 1930: During a visit to Cambridge, Albert Einstein stayed with Arthur Stanley Eddington and his sister Winifred. For Einstein it was an opportunity to be updated on cosmological matters. On Friday, 10 January 1930, Eddington had attended a lecture by Willem de Sitter, who, in a meeting of the Royal Astronomical Society, had talked about Hubble's observationally found linear relationship between the radial velocities of distant spiral nebulae and their distances (Hubble 1929). Neither de Sitter nor Eddington could provide a theoretical explanation, although both agreed that it had to be the consequence of some cosmological manifestation. The discussion of the two scientists was published in *The Observatory* (February 1930), whereupon George Lemaître sent them his publication of 1927. There he had shown that redshifts were the signature of an expanding universe, and that such redshifts depended linearly on the nebular distances (Lemaître 1927). Both, Eddington as well as de Sitter, immediately accepted Lemaître's theory of a dynamic universe; they published their opinion in March and May 1930 (Eddington 1930, de Sitter 1930a). For a history of these events see chapter 11 of Nussbaumer and Bieri (2009).

Eddington's publication contained another important message: Einstein's static universe of 1917 (Einstein 1917) was not stable. This was a strong argument for looking for dynamic cosmological solutions to Einstein's fundamental equations of general relativity. We do not know whether Eddington informed Einstein in writing about his death blow to his static world, no such correspondence has been found. However, we may be absolutely certain that Einstein learnt about it not later than June 1930. From the AEA 9-298 (Albert Einstein Archives) we know that on May 2nd Eddington invited Einstein to stay with him: "*My sister and I will be delighted if you and Mrs. Einstein will stay with us on this occasion. It would be very jolly to see you again. We should, of course, hope to have you with us for a few days so that you could see Cambridge at leisure*" (On May 2nd Eddington invites Einstein for May 5th, but he certainly meant June 5th). The visit of June 1930 was also documented by Eddington's sister who photographed the two scientists, and it is mentioned in Eddington's biography by Vibert Douglas (1956).

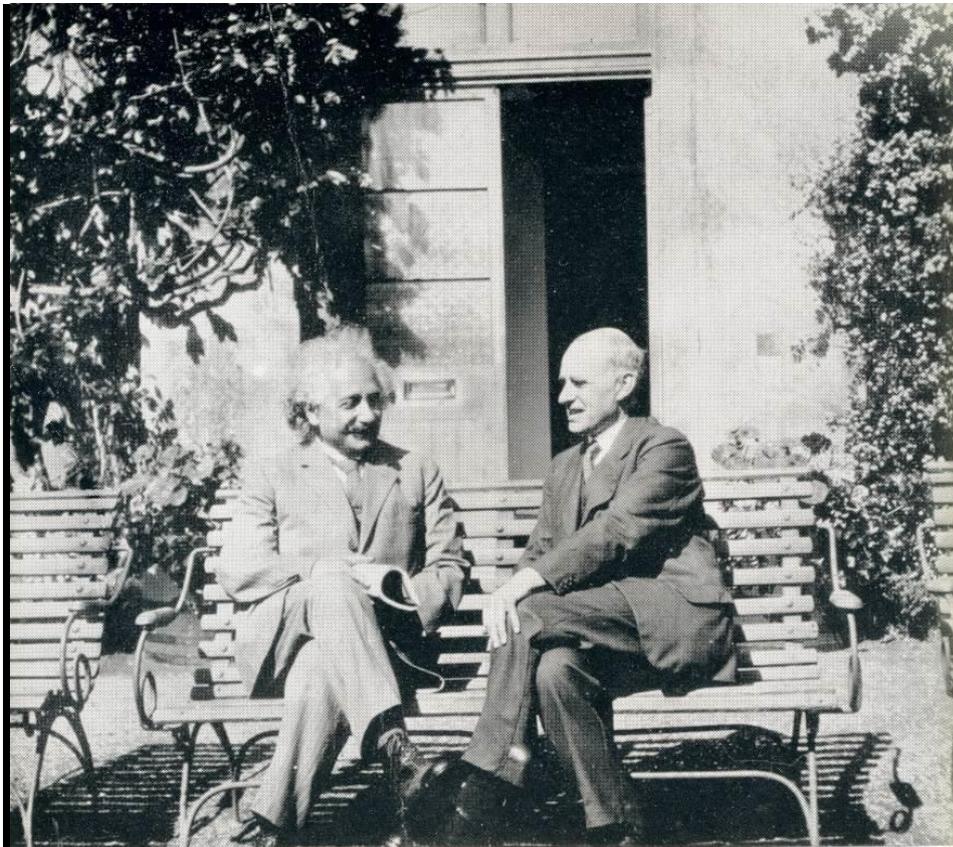

**A. Einstein and A.S. Eddington in front of Eddington's home.** Photographed June 1930 by Eddington's sister Winifred (Douglas 1956).

Thus, Einstein stayed with Eddington. It is unthinkable that on this occasion Eddington did not update Einstein on the latest evolution in cosmology: 1) Lemaître's discovery of the expanding universe, where Einstein would have to admit that he knew about the paper but did at the time not realise its significance, 2) Hubble's publication of crucial nebular observations and de Sitter's verification, 3) the adoption of Lemaître's theoretical model by de Sitter and Eddington, declaring Einstein's as well as de Sitter's models of 1917 as obsolete. But, for Einstein the most crucial point was probably Eddington's demonstration that his static universe of 1917 was unstable. This must have been the beginning of Einstein's conversion from his static to a dynamic universe. For the full story see Nussbaumer (2014) as well as O'Raifeartaigh and McCann (2014).

### Einstein's view of cosmology in 1928

Helge Kragh, in his forthcoming book *Masters of the Universe. Conversations with Cosmologists of the Past*, reports on a fictitious interview on 12 November 1928 in Einstein's home at Haberlandstrasse 5, Berlin (Kragh 2014). Talking cosmology, Einstein comments on de Sitter's empty universe of 1917: *I don't believe in de Sitter's solution; after all, the universe does contain matter, and a lot of it, and that's not its only weakness. I still believe that the best model is my original one, you know, all the matter in the universe is embedded in a closed and therefore finite space, the size of which can be calculated.* […] *the universe is finite in space and infinite in time. That's the punch line.* He also had something to say about Friedman's dynamic solutions (Friedman 1922): *"I should perhaps have given the paper more thought, but this is the kind of things that happens. In any case, from a physical point of view his idea of evolutionary solutions was unacceptable, and that's what matters"*. Concerning his scientific endeavours Kragh's confidant reports that Einstein didn't follow the literature on cosmology, yet: *"I haven't lost interest in the field, but for the last several years I have concentrated on extending the equations of general relativity from a theory of gravitation to one that describes also electromagnetism, and does it in a unified way that perhaps even describes elementary particles and not only fields"*.

Although fictitious, these "confessions" are likely to quite accurately define Einstein's cosmological position before he visited Eddington in June 1930. However, we now jump to the end of 1930. Einstein has become aware that his cosmology needs revision. He has arranged a visit to Pasadena, and we will finally learn from Einstein himself about his struggle with cosmology.

### Einstein's visit to Pasadena, from January 1 to the end of February 1931

It is still frequently claimed that Einstein abandoned his static model, when Hubble showed him his observations on the occasion of Einstein's first trip to Pasadena. This is purely fictional. Caltech had for many years tried to lure Einstein to Pasadena for a few weeks or months. On November 30, 1930 Einstein and his entourage left Berlin. After crossing the Atlantic and the channel of Panama the ship arrived in San Diego in the early morning of December 31. A detailed report about that journey and Einstein's conversion to a dynamic universe is given by Nussbaumer (2014).

When Einstein arrived in Pasadena, there was nothing new that Hubble could have taught him. Einstein knew about Hubble's publication on the nebular redshifts, although he always misspelt the name as Hubbel. Eddington had certainly informed him that de Sitter had checked Hubble's work and had arrived at the same observational conclusions. Thus, observationally the case was settled. It was now for the theoreticians to find the explanation.

On January 2, 1931 Einstein gave an interview to the *New York Times*. He implicitly admitted his helplessness in respect to the burning question: what could replace his static universe of 1917? He stated that his former attitude had become untenable. He told the journalist that observations by Hubble and Humason supported the assumption that the general structure of the universe was not static, and theoretical work by Tolman (Caltech) and Lemaître showed a view that fitted well into the theory of general relativity. Einstein did not volunteer a model of his own. This remained so during the full length of his two months stay in Pasadena. During this time he gave two interviews to the *New York Times* and two to the *Los Angeles Times*. Although Einstein's top priority in those days was the unified field theory, cosmology was mentioned in all four interviews. The tenor was the same in all of them. He knew that his original static universe was obsolete, but he could not warm up to any of the published alternatives. He brushed them aside, but offered no new solution.

### Richard C. Tolman (1881-1948), the cosmologist at Caltech

On January 3, 1931 we find a cryptic entry in Einstein's diary: „*Arbeiten im Institut. Zweifel an Richtigkeit von Tolmans Arbeit über kosmologisches Problem. Tolman hat aber Recht behalten*". (Work at the Institute. Doubts about the correctness of Tolman's work on the cosmological problem. But Tolman is right.) Let us briefly recall Tolman's cosmological contribution during those years. In 1929 and 1930 he burst onto the cosmological scene with half a dozen publications.

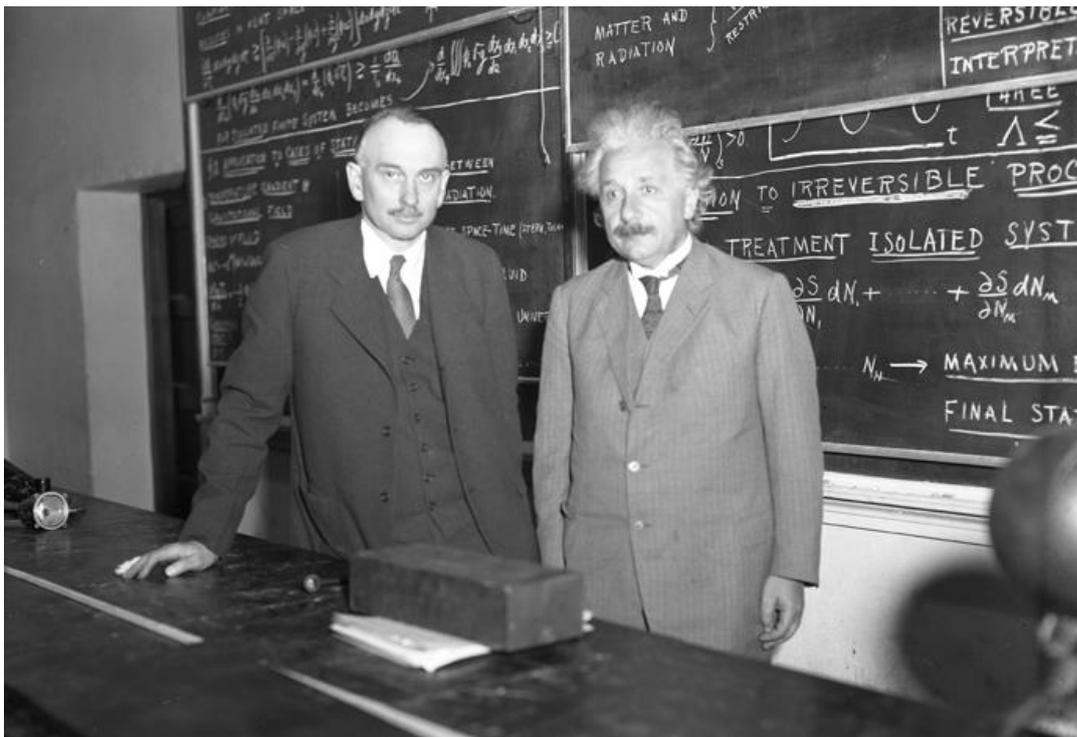

**Richard C. Tolman and Albert Einstein at the California Institute of Technology**. Published January 9, 1932, Los Angeles Times photographic archive, UCLA Library (Wikipedia).

The principal distinguishing factor of Tolman's contribution was the suggestion of annihilation of matter. Tolman was intrigued by de Sitter's model where test particles in the empty universe would disappear out of the observable universe. He had become aware of Robertson's dynamical line element (Nussbaumer and Bieri 2009, chapter 13), and he agreed that no static line element could account for the observed redshifts. This inspired him to the idea that transformation of matter into radiation was taking place throughout the universe. In that case the line element for an object at distance $r$ could not be static, but would be dynamic (Tolman 1930a):

$$ds^2 = -\frac{e^{2kt}}{\left(1 + \frac{4r^2}{R^2}\right)^2}\left(dx^2 + dy^2 + dz^2\right) + dt^2 \quad .$$

About $R$ and $k$ Tolman says: „*R is a constant having approximately the same significance as in the Einstein line element and k a constant directly related, on the one hand, to the average rate of transformation of matter into radiation and, on the other hand, to the known shift in wave-length with distance*". Dynamic did not imply that the universe had to be expanding or contracting.

Eddington criticised Tolman's approach (Eddington 1930, p. 671). However, after becoming aware of Tolman's suggestion, de Sitter considered that process as well, but arrived at the conclusion that annihilation could not be the cause of expansion; he reasoned that expansion was due to the cosmological constant (de Sitter 1930b).

Discussions on annihilation of particles had been going on ever since Eddington had tentatively suggested it as a possible source, compensating missing energy, when gravitational contraction was the only known energy for stellar radiation (Eddington 1917). In a last publication on this subject Tolman answered the criticism of Eddington and de Sitter, and he commented on the models of Friedman and Lemaître (Tolman 1930b). He emphasised the theoretical and observational obstacles that, in his mind, prevented an unequivocal choice of the appropriate line element. Anyhow, by that time there was general agreement that the line element had to represent a dynamic universe.

It is very likely that in June 1930 Eddington informed Einstein about Tolman's hypothesis of annihilation and that Einstein was looking forward to discuss these matters with Tolman. Indeed, according to Einstein's diary he debated cosmology with Tolman practically immediately after arriving in Pasadena.

### The document AEA 2-112

Document 2-112 of the AEA is a short draft "Zum kosmologischen Problem" (on the cosmological problem), handwritten by Einstein and assigned by AEA to the year 1931 (Einstein 1931a). The document AEA 2-112 can be seen on the internet http://alberteinstein.info/vufind1/Record/EAR000034354. From its obvious relation to Tolman's papers, its lack of references to Einstein's model of 1931 (Einstein 1931b), being written on American paper, and its likely relation to an entry in Einstein's diary on January 3, 1931, I date it to the beginning of January 1931.

At the beginning of the draft Einstein recalls his field equations of general relativity and that in 1917 he had introduced the cosmological constant λ (today generally designated by Λ) in order to generate a static universe. He now writes these modified equations as

$$\left(R_{ik} - \frac{1}{2}g_{ik}R\right) - \lambda g_{ik} = \kappa T_{ik} \quad (1) ,$$

(the numbering of the equations is the same as in Einstein's draft), and he continues:

"*Seither hat es sich so gut wie sicher herausgestellt, dass diese Lösung für die theoretische Erfassung des wirklichen Raumes nicht in Betracht kommt.*

*Einerseits geht nämlich aus den auf die nämlichen Gleichungen gegründeten Untersuchungen von ……. [space left open by Einstein] und von Tolman hervor, dass es auch sphärische Lösungen mit zeitlich veränderlichem Weltradius P gibt, und dass die von mir gegebene Lösung bezüglich der zeitlichen Aenderungen von P nicht stabil ist. Andererseits haben die überaus wichtigen Untersuchungen von Hubbel [sic] gezeigt, dass die extragalaktischen Nebel folgende Eigenschaften haben:*

*1) Sie sind in den Grenzen der Beobachtungs-Genauigkeit räumlich gleichmässig verteilt*

*2) Sie besitzen einen mit ihrer Distanz proportionalen Doppler-Effekt.*

(Since then it has turned out as practically certain that this solution cannot be considered a real solution for theoretically covering real space. On the one hand investigations by … [space left open by Einstein] and Tolman show that there exist spherical solutions with time-variable world-radius, and that my solution is not stable in respect to changes in time. On the other hand, the very important investigations by Hubbel [sic.] have shown that the extragalactic nebulae have the following properties:

1) Within the limits of observational accuracy they are homogenously distributed in space

2) They possess a Doppler-shift proportional to their distance.

For Einstein these statements define the essential cosmological facts known at the time when the draft was written. He acknowledges that his static model had been shown to be unstable and that there exist valid dynamical solutions to his fundamental equations. They have been found by Tolman and others, for whom he leaves an open space. He might have thought of Friedman's and Lemaître's solutions which he had snubbed some years ago. In addition there are Hubble's observations of a linear distance-redshift relationship.

Einstein then states that Tolman and de Sitter had shown that there were solutions to the modified fundamental equations (1), which did justice to the observations (*welche den Beobachtungen gerecht werden*) – he gives no citations. However, Einstein does not consider them as valid solutions of the cosmological problem. Because of the high values of the expansion rate of approximately 500 (km/s)/Mpc determined at the time these theories led to an age of the universe of approximately $10^{10}$ years. But in the early 1930s astronomers estimated stellar ages to be of the order of $10^{12}$ years or higher (see chapter 15 of Nussbaumer and Bieri (2009)). Thus an age of the universe of $10^{11}$ years or lower was deemed unacceptable.

Einstein then proposes a solution in which the density of the universe remains constant in time and which satisfies Hubble's observational results. "*Diese Lösung ist zwar in dem allgemeinen Schema Tolman's enthalten, scheint aber bisher nicht in Betracht gezogen worden zu sein*". (In fact, this solution is contained in Tolman's general scheme, but it seems that until now it has not been taken into consideration.)

Einstein then introduces a spatially Euclidean line element:

$$ds^2 = -e^{\alpha t}\left(dx_1^2 + dx_2^2 + dx_3^2\right) + c^2 dt^2 \qquad (2)$$

(In the draft the minus sign looks as if it had only been inserted in a second step.) – He states that the manifold is spatially Euclidean. He also points out that the distance between two points, when measured with a measuring rod, grows with time as $e^{\alpha t/2}$, and that Hubble's Doppler-shifts are obtained by assigning coordinates to the particles that remain constant in time. After a coordinate transformation, t'= t-τ, he re-writes the line element as

$$ds^2 = e^{\alpha t'}\left(dx_1^{'2} + dx_2^{'2} + dx_3^{'2}\right) + c^2 dt^{'2}\,.$$

(He forgets the minus sign.) In his matter-tensor he neglects pressure terms and writes:

$$T^{ik} = \rho u^i u^k \quad \left(u^i = \frac{dx^i}{ds}\right)$$

or $T_{ik} = \rho u^\sigma u^\tau g_{\sigma i} g_{\tau k}$  (3)

with $u^1 = u^2 = u^3 = 0; \quad u^4 = \frac{1}{c}$ .

The fundamental equations (1) provide him with two relations between α, ρ, and λ, which he writes as

$$\frac{9}{4}\alpha^2 + \lambda c^2 = 0$$

$$\frac{3}{4}\alpha^2 - \lambda c^2 = \kappa \rho c^2$$

from which he derives a relation between α, κ, and ρ:

$$\alpha^2 = \frac{\kappa c^2}{3}\rho \qquad (4)\,.$$

Einstein comments that the density remains constant and determines the expansion except for its sign. (The factor 9/4 cannot be read with absolute certainty, according to information from AEA it might also be 5/4. However, only the value 9/4 leads to the numerical coefficient of relation (4).)

Einstein terminates the draft with the following statements: "*Betrachtet man ein durch physische Massstäbe begrenztes Volumen, so wandern ausausgesetzt materielle Teilchen aus demselben hinaus. Damit die Dichte konstant bleibe, müssen immer neue Massenteilchen in dem Volumen aus dem Raume entstehen. Der Erhaltungssatz bleibt dadurch gewahrt, dass bei Setzung des λ-Gliedes der Raum selbst nicht energetisch leer ist; seine Geltung wird bekanntlich durch die Gleichungen (1) gewährleistet*". (If one considers a volume, limited by a physical measuring rod, then material particles will incessantly leave the volume. In order to keep the density constant, new particles will permanently have to come into existence in this volume out of space. The conservation law is preserved, in that by accepting the λ-term, space is energetically not empty; as is well known this fact is assured by the equations (1).)

We have seen above that a line element with a time-dependent factor $e^{2kt}$ had already been introduced by Tolman in 1930, where $k$ served as a measure for the annihilation of matter and also provided the observed redshifts. Einstein now assigns to α the role of a creation source. He changes his former assumption of a static universe to a dynamic, expanding universe, of constant density ρ. The cosmological constant, λ, is no longer an embarrassing stopgap, but assumes a physical quality: it provides the energy necessary for creating new particles. To Einstein this solution must have come as a very satisfying way out of the cosmological dilemma. It turned his static model into a stationary, dynamic equilibrium of unlimited lifetime, the awkward age problem had disappeared, and physically it was not less implausible than Lemaître's, Eddington's, and de Sitter's λ-driven expansion.

However, Einstein's relief was probably very short lived. The draft shows that he corrected the original version and found a numerical error. In the upper one of the two expressions that led to relation (4) the coefficient in front of $\alpha^2$ was changed from 9/4 to -3/4. The resulting two equations

$$-\frac{3}{4}\alpha^2 + \lambda c^2 = 0$$

$$\frac{3}{4}\alpha^2 - \lambda c^2 = \kappa\rho c^2$$

lead directly to $\rho=0$. Thus, Einstein had basically fallen back into de Sitter's empty universe. He did not bother to correct his relation (4); obviously the attempt had collapsed.

**Document AEA 2-112 from the Albert Einstein Archive**. Einstein thought to have found a relation between $\rho$, the density of the universe, and the creation parameter $\alpha$.

We do not know whether Einstein found the error himself. When I saw the correction to the first term and its devastating effect on the conclusion, it reminded me immediately of Einstein's diary with his entry on January 3, 1931: "Doubt about correctness of Tolman's work on the cosmological problem. But Tolman was right". Let us look at Einstein's original, German formulation. Einstein's draft AEA 2-112 carries the title: "*Zum kosmologischen Problem*". Compare this to the entry in his diary on January 3: "*Zweifel an Richtigkeit von Tolmans Arbeit über kosmologisches Problem. Tolman hat aber Recht behalten*". Although this coincidence does prove the case for placing Einstein's draft squarely to the very beginning of January 1931, it presents an additional argument.

Tolman was one of the first to greet Einstein at Pasadena and they immediately engaged in cosmological discussions. Whether already primed by Eddington, or initialised by Tolman, Einstein was probably taken with the novel idea of annihilation: why not transform it into creation! His remark that his approach was already contained in Tolman's general scheme, but until then had not been taken into consideration, suggests that his draft grew out of a glance through Tolmans papers or in the aftermath of a personal conversation.

*Concluding remarks*

If my dating of (AEA 2-112) is correct, then Einstein's draft belongs to the period between his visit to Eddington in June 1930 and the publication of his dynamic Friedman-Einstein model in April 1931 (Einstein 1931b). During those months his scientific priority was the unified-field-theory. However, after the visit to Eddington he was confronted with the latter's demonstration that the static universe, proposed in his ground-breaking publication of 1917, was unstable. This forced him to redefine his cosmology. We saw from the interviews given to the *New York Times* and the *Los Angeles Times* during his two months stay in Pasadena that the alternatives offered by Friedman and Lemaître did not find his approval (Nussbaumer 2014). Whereas Friedman presented purely mathematical models, Lemaître had already anticipated Hubble's observational redshift-distance relationship in 1927. The combination of Lemaître's and Hubble's contributions had convinced Eddington and de Sitter, two acknowledged authorities in astrophysics, that our universe was in expansion. But Einstein stubbornly refused to accept Lemaître's view. The main reason was probably that in Einstein's mind the expanding model pointed to a beginning of our present universe, which was far too recent to be compatible with the ages of the stars. In this respect Einstein was mistaken. Lemaître, as well as Eddington, offered solutions that could well accommodate high stellar ages. When Eddington showed that Einstein's universe was unstable, he suggested that a pseudo-static universe might have been the original status, out of which expansion was born. Also Lemaître's universe of 1927 increased asymptotically since time -∞. After Lemaître had in 1931 suggested the beginning of the universe out of a single atom, containing all the matter of the universe, he saw the balance between gravitational contraction and acceleration by the cosmological constant regulating the further history of the universe. This also gave plenty of time for the formation and evolution of stars.

The first version of the draft AEA 2-112 seemed to promise a way out of the dilemma: a stationary universe in dynamical equilibrium, with no beginning and no end, matter being created out of the vacuum, showing properties of expansion. Einstein was not the first to suggest spontaneous creation of matter. In 1928, J.H. Jeans, on page 352 of his book *Astronomy and Cosmogony* [Jeans 1928, last paragraph of chapter 13], had conjectured: *the centres of the nebulae are of the nature of «singular points,» at which matter is poured into our universe from some other, and entirely extraneous, spatial dimensions, so that, to a denizen of our universe, they appear as points at which matter is being continually created.* Eddington probably knew that book and might have told Einstein about Jeans' speculation. But it is more likely that Einstein conceived the idea of creation after having become acquainted with Tolman's annihilation hypothesis. The model outlined in his draft fitted his quest for a never changing universe very well.

Although AEA 2-112 did not influence the course of cosmology, it is an intriguing document about Einstein's thinking. He was doubtless inspired by Tolman's ideas of annihilation. Tolman's universe could in principle have been a "thinning out" version of Einstein's model of 1917, where particles were disappearing, converting into radiative energy. In his draft, which I place tentatively to the beginning of January 1931, Einstein transformed this idea into a stationary, dynamical universe. Local conditions would remain the same. New particles were permanently being created inside a given volume, and the same number of particles would leave that volume, thus contributing to the expansion of the universe. He kept his cosmological constant, $\lambda$, it provided the energy for creating the new particles.

Similar ideas would be developed 17 years later by Hermann Bondi, Tommy Gold, and Fred Hoyle (Bondi and Gold 1948; Hoyle 1948). The decisive difference in Gold's approach was his treatment of $\lambda$. He replaced $\lambda$ by a creation vector $C_\mu$, from which he obtained a tensor field $C_{\mu\nu}$. With this $C_{\mu\nu}$ he replaced the $\lambda$-term on the left-hand side of the fundamental equations (1). However, contrary to $\lambda$, the only non-vanishing components in $C_{\mu\nu}$ are the spatial contributions ($\mu, \nu = 1, 2, 3$ in Hoyle's notation). For the right-hand side of the fundamental equations he kept as only non-vanishing component $T_{oo} = \rho c^4$ (in Hoyle's notation). With his relation (3) Einstein had done the same. Hoyle then comments: "The $C_{\mu\nu}$ term […] plays a role similar to that of the cosmological constant in the de Sitter model, with the important difference, however, that there is no contribution from the $C_{00}$ component. As we shall see, this difference enables a universe, formally similar to the de Sitter model, to be obtained, but in which $\rho$ is non-zero".

Einstein's draft does not tell us why he abandoned his attempt. But if his diary of 3 January 1931 really holds the clue ("but Tolman is right") then it must have been the outcome of discussions with Tolman. They might well have gone beyond the question of an error in the calculations. There are no indications that Einstein pursued the idea of a stationary universe in a dynamical equilibrium, creating matter out of the vacuum, any further. We should not forget: at that time cosmology to him was of secondary importance, his mind was preoccupied with the unified field theory.

We should not judge Einstein's draft on the same level as the polished steady-state theory developed by Bondi, Gold, and Hoyle. Einstein didn't spend much time on his sudden inspiration about "creation out of empty space". It had been an idea which grew out of Tolman's hypothesis of annihilation, and to which Einstein simply tried to give a different twist. However, all four scientists were inspired by the same basic philosophical hypothesis: there may be a cosmic energy source – Einstein had given it a name with his cosmological constant – supporting the creation of new particles. It could drive the expansion of the universe and thus explain the observed redshifts of remote extragalactic nebulae. That process might be active within an eternal universe, and thus accommodate the estimated ages of the stars, highly exaggerated up to the middle of the 1930s. The original vision of Bondi, Gold, and Hoyle, launched in 1948, was only terminated in 1965 with the discovery of the 3-degree background radiation (on this subject see also Kragh 1999). Einstein's attempt was probably shipwrecked within one or two days. It was a shot from the hip that misfired.

Even before I came across the document AEA 2-112 Cormac O'Raifeartaigh had, unknown to me, found it as well. He and his collaborators are preparing a discussion and translation of Einstein's aborted attempt; their work has now appeared (O'Raifeartaigh et al. 2014).

I thank Ms. Barbara Wolff at the Albert Einstein Archives in Jerusalem, for inspecting the original document AEA 2-112, and Lydia Bieri for discussions. — Hilmar Duerbeck would doubtless have enjoyed to comment on Einstein's draft, tucked away unnoticed in the drawer of an archive. I dedicate this contribution to Hilmar's memory.